\documentstyle[pra,aps,amsfonts,preprint]{revtex}
\begin{document}
\draft
\title{Comment on ``Experimental Nonlocality Proof of Quantum Teleportation and
Entanglement Swapping''} 
\author{Luiz Carlos Ryff}
\address{Instituto de F\'{\i}sica, 
Universidade Federal do Rio de 
Janeiro, Caixa Postal 68528, Rio de Janeiro, RJ 21941-972, Brazil} 
\date{\today}
\maketitle
 In a recent paper in which they demonstrate quantum nonlocality
for photons that never interacted \cite{1}, Jennewein, Weihs, Pan, and
Zeilinger mention a seemingly paradoxical situation that arises when Alice's
Bell-state analysis is delayed long after Bob's measurements. According to
the authors, ``This seems paradoxical because Alice's measurement projects
photons 0 and 3 into an entangled state after they have been measured,'' in
other words, ``This means that the physical interpretation of his (Bob's)
results depends on Alice's later decision.'' Still quoting from their paper:
``Therefore, this result indicates that the time ordering of the detection
events has no influence on the results and strengthens the argument of A.
Peres \cite{2}: \textit{this paradox does not arise if the correctness of
quantum mechanics is firmly believed.''}

The authors are correct about the time order of the detection events.
Actually, Einstein, Podolsky, and Rosen (EPR) correlations, teleportation,
and entanglement swapping are different aspects of the very same phenomenon
when viewed from different Lorentz frames that move relative to each other \cite{3}. 
On the other hand, they are incorrect when they conclude that the
physical interpretation of Bob's results depends on Alice's later decision,
and Peres' statement, to which they subscribe, does little to explain why
there is no paradox. The seeming paradox arises because, apparently, a
measurement in the future projects the photons in an entangled state in the
past \cite{4} . If this were true, this would imply that not only local realism,
as shown by Bell \cite{5}, but also nonlocal realism cannot mimic quantum
mechanics, since there can manifestly be no causal connection between the
events observed by Bob (even accepting the possibility of superluminal
interaction). In particular, Bohm's interpretation \cite{6} would be condemned.
Therefore, it is important to show why there is no paradox.

Naturally, firm belief is not the most appropriate way to try to solve a
paradox. Actually, the reason why such a paradox does not occur is quite
simple and has already been explained \cite{7}. The point is that whenever Bob
performs his measurement first, photons 1 and 2 are projected into
well-defined polarization states. As a consequence, Alice will perform an
experiment similar to the one performed by Hong, Ou, and Mandel \cite{8}, but
in which the photons that arrive together at the beam splitter can have
different polarizations. Since the probability of coincident detection
depends on the relative polarization of the photons, by selecting the events
in which the detectors click in coincidence, Alice obtains a subset that
behaves as if it consisted of entangled pairs of distant particles.
Naturally, there is nothing paradoxical in this result. The point is that
for Alice the probability of coincident detection will depend on Bob's
experimental outcome. Therefore, no influence of future actions on past
events needs to be invoked. Very probably, the authors were not aware of
this simple explanation, or else they would realize that there is no reason
to perform the experiment discussed by Peres, since it adds nothing to what
is already known. The photon-photon entanglement in type-II spontaneous
parametric down-conversion and the behavior of photons that arrive together
at a beam splitter have already been experimentally verified. Naturally, the
authors have performed an important and interesting experiment, but their
extension to the situation in which Alice's measurement occurs after Bob's
is unnecessary and their interpretation misleading.

{\bf ACKNOWLEDGMENT}

Financial support was provided by Brazilian agency CNPq-Institutos do Mil\^enio-
Informa\c c\~ao Qu\^antica


\begin{references}
\bibitem{4} 
According to Peres, ``quantum effects mimic not only
instantaneous action-at-a-distance but also influence of future actions on
past events'' (ref. [2]).
\end{references}
\end{document}